\documentclass[iop]{emulateapj} 

\usepackage{graphicx}
\usepackage{epstopdf}

\usepackage{gensymb}

\usepackage{savesym}
\savesymbol{tablenum}
\usepackage{siunitx}
\restoresymbol{SIX}{tablenum}

\let\oldenddeluxetable\enddeluxetable
\let\olddeluxetable\deluxetable
\makeatletter
\renewenvironment{deluxetable}[1]{
\olddeluxetable{[#1]}
\def\pt@format{#1}%
}{\oldenddeluxetable}
\makeatother

\usepackage{xfrac}
\usepackage{hyperref}
\hypersetup{
	colorlinks=true,
	citecolor=blue,
	linkcolor=blue
}

\renewcommand{\figurename}{Figure}
\renewcommand{\tablename}{Table}

\slugcomment{Submitted to Astrophysical Journal}

\shorttitle{\emph{NuSTAR} and \emph{XMM-Newton} observations of 1E1743.1-2843}
\shortauthors{Lotti et al.}

\begin{document}

\title{\emph{NuSTAR} and \emph{XMM-Newton} observations of \object{1E1743.1-2843}:
indications of a neutron star LMXB nature of the compact object}

\author{Simone Lotti\altaffilmark{1}, Lorenzo Natalucci\altaffilmark{1}, Kaya Mori\altaffilmark{2},
Frederick K. Baganoff\altaffilmark{3}, 
Steven E. Boggs\altaffilmark{4}, 
Finn E. Christensen\altaffilmark{5}, 
William W. Craig\altaffilmark{4,3}
Charles J. Hailey\altaffilmark{3},
Fiona A. Harrison\altaffilmark{6},
Jaesub Hong\altaffilmark{7}, 
Roman A. Krivonos\altaffilmark{4,8}, 
Farid Rahoui\altaffilmark{9,10},
Daniel Stern\altaffilmark{11},
John A. Tomsick\altaffilmark{4},
Shuo Zhang\altaffilmark{3},
William W. Zhang\altaffilmark{12}
}
\altaffiltext{1}{\footnotesize{INAF-IAPS Roma, Via fosso del cavaliere 100, Rome 00133, Italy; e-mail: \email{simone.lotti@iaps.inaf.it} }}
\altaffiltext{2}{\footnotesize{Columbia Astrophysics Laboratory, Columbia University, New York, NY 10027, USA}}
\altaffiltext{3}{\footnotesize{MIT Kavli Institute for Astrophysics and Space Research, Cambridge, MA 02139, USA}}
\altaffiltext{4}{\footnotesize{Space Sciences Laboratory, 7 Gauss Way, University of California, Berkeley, CA 94720-7450, USA}}
\altaffiltext{5}{\footnotesize{DTU Space, National Space Institute, Elektrovej 327, DK-2800 Lyngby, Denmark}}
\altaffiltext{6}{\footnotesize{Cahill Center for Astronomy and Astrophysics, Caltech, Pasadena, CA 91125, USA}}
\altaffiltext{7}{\footnotesize{Harvard-Smithsonian Center for Astrophysics, 60 Garden Street, Cambridge, MA 02138, USA}}
\altaffiltext{8}{\footnotesize{Space Research Institute, Russian Academy of Sciences, Profsoyuznaya 84/32, 117997 Moscow, Russia}}
\altaffiltext{9}{\footnotesize{Department of Astronomy, Harvard University, 60 Garden Street, Cambridge, MA 02138, USA}}
\altaffiltext{10}{\footnotesize{European Southern Observatory, Karl Schwarzschild-Strasse 2, 85748 Garching bei M\"unchen, Germany}}
\altaffiltext{11}{\footnotesize{Jet Propulsion Laboratory, California Institute of Technology, Pasadena, CA 91109, USA}}
\altaffiltext{12}{\footnotesize{NASA Goddard Space Flight Center, Greenbelt, MD 20771, USA}}

\begin{abstract}
We report on the results of \emph{NuSTAR} and \emph{XMM-Newton} observations of the persistent X-ray source \object{1E1743.1-2843}, located in the Galactic Center region. The source was observed between September and October 2012 by \emph{NuSTAR} and \emph{XMM-Newton}, providing almost simultaneous observations in the hard and soft X-ray bands. The high X-ray luminosity points to the presence of an accreting compact object. We analyze the possibilities of this accreting compact object being either a neutron star (NS) or a black hole, and conclude that the joint \emph{XMM-Newton} and \emph{NuSTAR} spectrum from 0.3 to 40 $\mathrm{keV}$ fits to a black body spectrum with $kT\sim1.8~\mathrm{keV}$ emitted from a hot spot or an equatorial strip on a neutron star surface. This spectrum is thermally Comptonized by electrons with $kT_{e}\sim4.6~\mathrm{keV}$. 
Accepting this neutron star hypothesis, we probe the Low Mass (LMXB) or High Mass (HMXB) X-ray Binary nature of the source. While the lack of Type-I bursts can be explained in the LMXB scenario, the absence of pulsations in the \hbox{2 mHz - 49 Hz} frequency range, the lack of eclipses and of an IR companion, and the lack of a $K_{\alpha}$ line from neutral or moderately ionized iron strongly disfavor interpreting this source as a HMXB. We therefore conclude that \object{1E1743.1-2843} is most likely a NS-LMXB located beyond the Galactic Center. There is weak statistical evidence for a soft X-ray excess possibly indicating thermal emission from an accretion disk. However, the disk normalization remains unconstrained due to the high hydrogen column density ($N_{H}\sim1.6\times10^{23}~\mathrm{cm^{-2}}$).

\end{abstract}


\section{Introduction}

The X-ray source \object{1E1743.1-2843} was discovered during the first soft X-ray imaging observation of the Galactic Center, performed by the \emph{Einstein Observatory} \citep{watson1981}, and has been detected in all the subsequent observations performed by X-ray satellites with imaging capabilities above $2~\mathrm{keV}$ \citep{kaway1988,sunyaev1991,pavlinsky1994,lu1996,cremonesi1999, chandra, bird, porquet2003}. Its position has been determined with \emph{Chandra} to be $\alpha_{J2000}~=~17^{h}~46^{m}~21.09^{s}$, $\delta_{J2000}~=-28^{\degree}43'~42.67''$ with a reported 0.21'' $1~\sigma$ accuracy \citep{chandrapos}. Because of its high column absorption ($N_{H}=1.3\pm0.1\times10^{23}~\mathrm{cm^{-2}}$, see \citealp{cremonesi1999}), the source is likely in the Galactic Center ($d=7.9\pm0.3~\mathrm{kpc}$, \citealp{mcnamara2000}) or beyond, while the orbital inclination is smaller than $70^{\degree}$ \citep{cowley1983}. The analysis performed by \citet{porquet2003} detected no pulsations or quasi-periodic oscillations (QPO) in the $2.4~\mathrm{mHz}-2.5
 ~\mathrm{Hz}$ frequency range using EPIC-MOS and PN fullframe mode (time resolution 2.6 s and 200 ms, respectively). However, since many X-Ray Binaries (XRB) present quasi-periodic variations above $2.5~\mathrm{Hz}$, the \emph{XMM-Newton} data were not suitable to probe the millisecond pulsations that could indicate a Low Mass X-ray Binary (LMXB) nature. In their analysis, Porquet tested several single-component spectral models (i.e., absorbed power-law, absorbed black body, absorbed disk-black body) but could not distinguish between these models due to the narrow bandpass from 2 to 10 $\mathrm{keV}$. In this regard the \emph{NuSTAR} capabilities to perform high resolution broadband spectroscopy allow us to probe into the nature of the high-energy emission from this source more deeply.

The presence of an accreting object is required to explain the unabsorbed source luminosity, which is of the order of $L_{2-10~\mathrm{keV}}\sim10^{36}~d^{2}_{10~\mathrm{kpc}}~\mathrm{erg~s^{-1}}$, where $d_{10~\mathrm{kpc}}$ is the source distance expressed in units of 10 kpc, while the absence of periodic oscillations and eclipses favors a scenario in which a compact object (either a neutron star or a black hole) accretes matter from a low mass companion (LMXB systems). LMXBs in this luminosity range containing a neutron star are usually characterized by thermonuclear flashes of the accreted matter that ignites on the neutron star surface (Type 1 X-ray bursts), but these bursts have never been observed for \object{1E1743.1-2843} in 20 years of X-ray observations. This has led so far to the conclusion that (1) the accretion rate is high enough to allow stable burning of the accreted material, which would imply a $\dot{M}$ value comparable to the Eddington limit \citep{bursters}, and thus distance greater than $8~\mathrm{kpc}$, (2) the bursts are suppressed by the presence of intense magnetic fields (at least $10^{9}~\mathrm{G}$), (3) the accreting compact object is a black hole \citep{porquet2003}. The presence of strong magnetic fields ($>10^{12}~\mathrm{G}$), however, should be accompanied by cyclotron absorption features and pulsations, neither of which have been observed in \object{1E1743.1-2843}. The source showed marginal variability on month timescales in the $20-40~\mathrm{keV}$ range \citep{delsanto2006}, some variability on hour timescales (10-20\%) in the $1.3-10~\mathrm{keV}$ energy range \citep{cremonesi1999}, and less than 18\% of variability between $10^{-4}$ and 2.5 Hz in the $2-10~\mathrm{keV}$ energy range \citep{porquet2003}.

We present here the results of four observations of \object{1E1743.1-2843} performed with the \emph{NuSTAR} and \emph{XMM-Newton} satellites in September-October 2012. The two satellites provided almost simultaneous broad-band X-ray spectroscopy from 0.1 to 79 $\mathrm{keV}$ and, thanks to the \emph{NuSTAR} timing capabilities, we can also probe pulsations down to 2 milli-second timescales, providing unprecedented opportunities to investigate this source. 
In section~\ref{sec:obs} we describe the observations, in section~\ref{sec:analysis} we discuss the timing and spectral analysis. Our results and discussion are presented in section \ref{sec:results}.

\section{Observations}
 \label{sec:obs} 


\begin{table*}
\begin{center}
\scriptsize
\caption{\protect\emph{NuSTAR} and \protect\emph{XMM-Newton} observations of \protect\object{1E1743.1-2843}.}
\label{tblobs}
\begin{tabular}{cccccc}
\tableline\\ [-2.0ex]
Obs.Id & Revolution & Satellite & Start (UTC) & End (UTC) & Exposure time (s) \\
\tableline\\ [-2.0ex]
40010005001 & - & \emph{NuSTAR} & 2012-10-15 13:31:07 & 2012-10-16 05:41:07 & 25993 \\
0694640401* & 2332 & \emph{XMM-Newton} & 2012-09-02 19:33:02 & 2012-09-03 11:20:55 & 24830 \\
0694640501 & 2334 & \emph{XMM-Newton} & 2012-09-05 21:16:14 & 2012-09-06 09:46:55 & 29581 \\
0694641201 & 2344 & \emph{XMM-Newton} & 2012-09-26 06:17:37 & 2012-09-26 17:24:59 & 35540 \\
\tableline\\ [-2.0ex]
\end{tabular}
\tablecomments{
$^*$ Data from obs. 0694640401 were not used in the analysis.}
 \end{center}
 \end{table*}

\begin{figure}
\includegraphics[width=\linewidth]{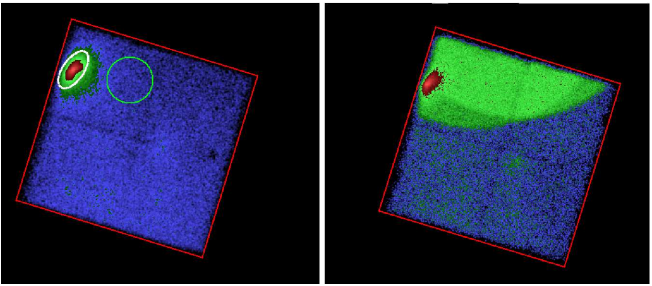}
\caption{Images from \emph{NuSTAR} focal plane modules in the 3-79 keV energy band, FPMA (left) and FPMB (right), the target source is $\sim7'$ off-axis during the observation. FPMB is highly contaminated by stray light, and moreover the source is cut in half by two different stray light zones, therefore only data from FPMA were used in the analysis.\label{straylight}}
\end{figure}

All the observations were performed between September and October 2012. \emph{NuSTAR} observed \object{1E1743.1-2843} during the a so-called ``mini-survey'' of the Galactic Center, which took place shortly after the \emph{NuSTAR} in-orbit checkout. The observation was performed on October 15th 2012, and the total exposure time was $26~\mathrm{ks}$. The source was $\sim7'$ off-axis in the \emph{NuSTAR} observation and appears distorted due to the asymmetric PSF shape at a large off-axis angle \citep{nustarcalib}, as can be seen in \figurename~\ref{straylight}.

\emph{NuSTAR} is the first X-ray satellite with multilayer hard X-ray optics and is operational in the energy range $3-79~\mathrm{keV}$ \citep{Harrison2013}. The mission carries two identical telescopes with grazing incidence optics, each one focusing on separate detector modules, Focal Plane Modules A and B (FPMA, FPMB), at a distance of 10~m. These CdZnTe detectors have a total Field of View (FoV) of $13'\times13'$ \citep{Harrison2013}. The telescope Point Spread Function (PSF) has an 18'' FWHM with extended tails resulting in a Half-Power-Diameter of 58'' \citep{Harrison2013}.

To improve the low-energy sampling of the source spectrum we looked for other high-energy observations of \object{1E1743.1-2843}, performed approximately during the same period. We adopted three \emph{XMM-Newton} observations performed in imaging mode during September 2012, for a total exposure time of $90~\mathrm{ks}$. Observation 0694640401 however, is strongly contaminated by solar emission, we therefore decided to not use it, reducing the useful exposure time to $65~\mathrm{ks}$. 
The \emph{XMM-Newton} EPIC-pn camera provides data with nominal accuracy in the $0.3-10~\mathrm{keV}$ energy band \citep{xmmguide}, providing a good overlap with the \emph{NuSTAR} data, thereby minimizing possible bias in the spectral modeling.

The \emph{NuSTAR} observations are not simultaneous with \emph{XMM-Newton}, but due to the small difference in time (few weeks, see~\tablename~\ref{tblobs}), and since \object{1E1743.1-2843} does not usually exhibit substantial spectral variability \citep{cremonesi1999}, we jointly fit the two datasets with a cross-normalization factor to account for any flux variation. During the observations the difference of source fluxes measured by the two instruments in the overlapping energy band ($3-10~\mathrm{keV}$) was below 1.3\%, indicating a good compatibility of the datasets.

\section{Data analysis}
 \label{sec:analysis} 
 
\subsection{\emph{NuSTAR}} We analyzed the \emph{NuSTAR} data set (obsID \dataset{40010005001}) using the \emph{NuSTAR} Data Analysis Software (NuSTARDAS) version 1.3.1 (9 December 2013), HEASOFT 6.15.1, and the most updated calibration files and responses. The software applies offset correction factors to the energy response to account for the movement of the mast that causes a varying position of the focal spots on the detector planes. For the data from each of the two modules, the pipeline produces an image, spectrum and deadtime corrected lightcurve. For each \emph{NuSTAR} observation, the source and background subtraction regions must be carefully evaluated due to the possible presence of contaminating sources outside the FoV that induce stray light patterns on the detectors. 
Unfortunately the \emph{NuSTAR} detectors are not entirely shielded from unfocused X-rays, and this stray light can be significant if there are bright X-ray sources within $\sim2-3$ degrees from the pointing direction \citep{stray1, stray2, stray3}. In this observation FPMB is highly contaminated by two different stray light patterns, as can be seen in \figurename~\ref{straylight}, and furthermore the source focal spot straddles the two different patterns. Due to the complexity of separating the source emission from the stray light we decided to discard all the data from FPMB.

\begin{figure}[ht] 
 \centering
 \includegraphics[width=\linewidth]{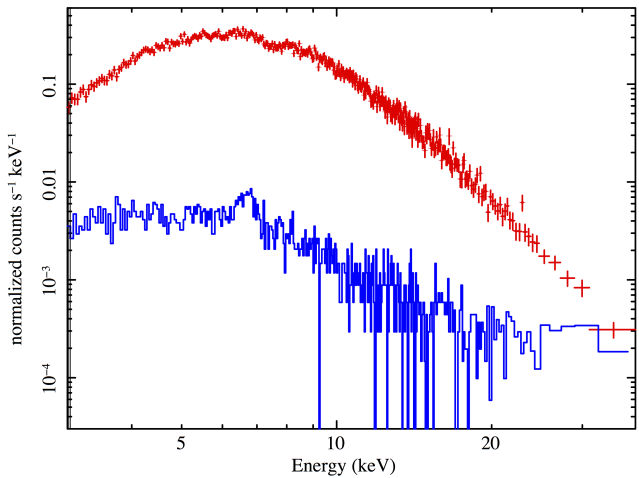} 
 \vspace{4ex}
 \caption{\label{nuspec}Spectrum acquired from the \emph{NuSTAR} FPMA extraction region. The red crosses indicate the source spectrum rebinned to have at least $5\sigma$ significance and 30 counts for each bin. The blue line is the background spectrum.}
\end{figure}

The source spectra are shown in \figurename~\ref{nuspec}, and were obtained by extracting photons in an elliptical region of $95''\times46''$ semi axis, rotated by 35 degrees clockwise relative to north and centered on the source centroid, and subtracting count rate measured in a nearby circular background region of radius 114'' (see \figurename~\ref{straylight}). The different area normalizations were taken into account in the background subtraction. The dead layer thickness of the two modules varies depending on the location on the detector and, since the pipeline could fail to correct the response matrix for this effect to a suitable accuracy for high off-axis angles, we decided to ignore the data below $5~\mathrm{keV}$. Also, due to imperfect background subtraction, the data above $40~\mathrm{keV}$ were ignored. All data bins were grouped to reach at least 30 counts. The total source count rate in this energy range is $2.10\pm0.01~\mathrm{cts/s}$.

\subsection{\emph{XMM-Newton}} \label{xmm} 
\begin{figure*}[ht]
 \centering
 \includegraphics[width=\linewidth]{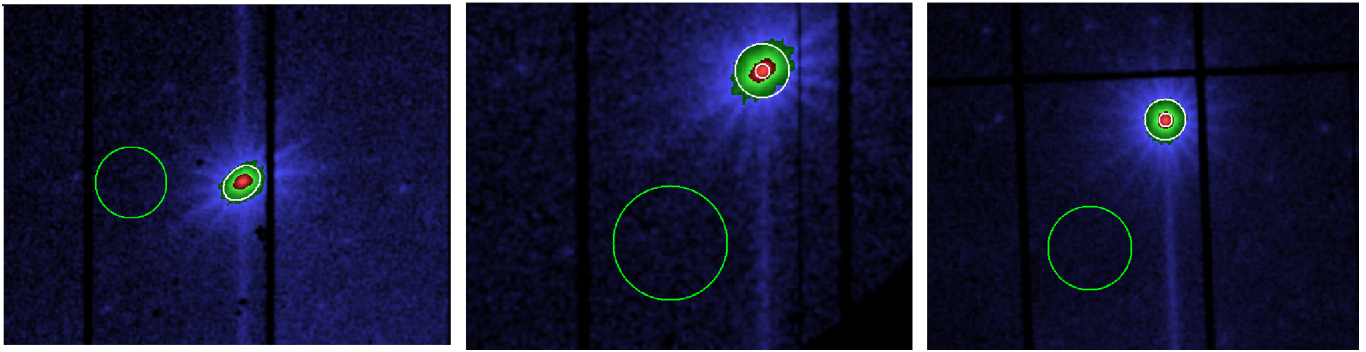} 
 \vspace{4ex}
\caption{\label{xmmregion}Images of \protect\object{1E1743.1-2843} from \protect\emph{XMM-Newton} EPIC-pn camera. White ellipses show the annular extraction region, and green circles show the background regions. The bright central zone of the source was excluded due to pileup in the central and rightmost panels, while the pileup threshold was not exceeded in the leftmost one (thus no inner ellipse was plotted on the image).}
\end{figure*}

\begin{figure*}[ht] 
 \begin{minipage}[b]{0.5\linewidth}
 \centering
 \includegraphics[width=1\linewidth]{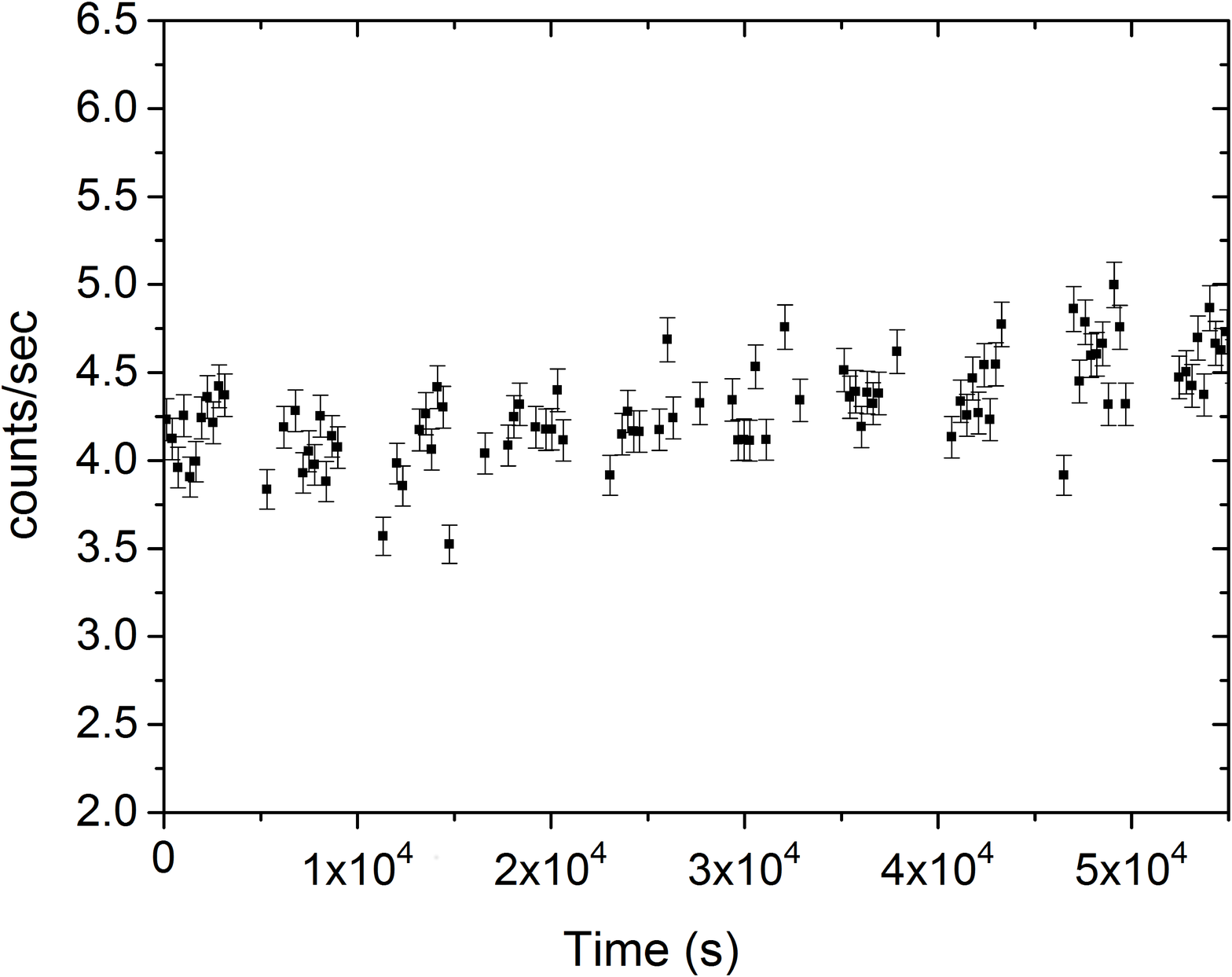} 
 \vspace{4ex}
 \end{minipage}
 \begin{minipage}[b]{0.5\linewidth}
 \centering
 \includegraphics[width=1\linewidth]{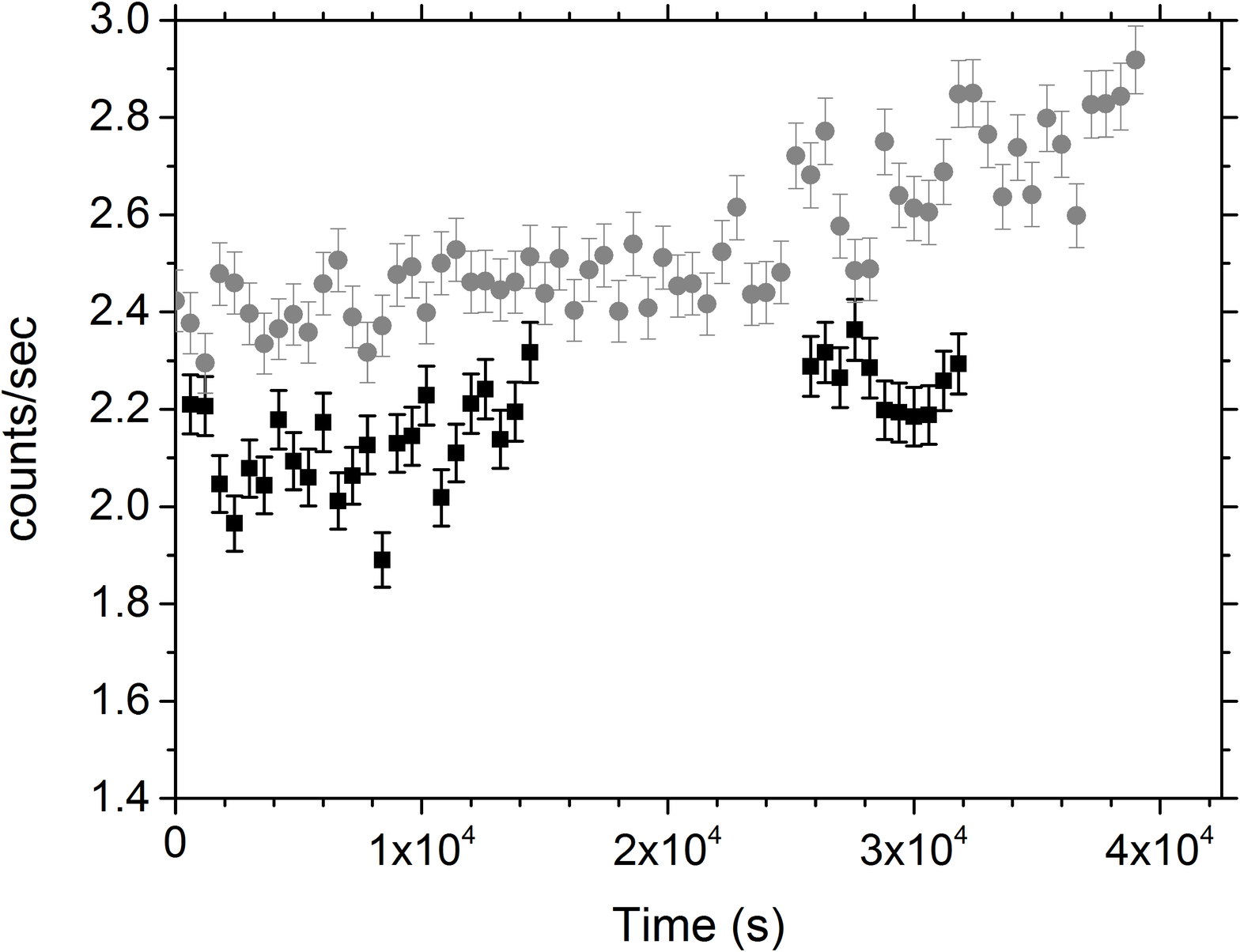} 
 \vspace{4ex}
 \end{minipage}
 \caption{ \label{lc}Light curves of 1E1743.1-2843 acquired from the \emph{NuSTAR} FPMA in the $3-80~\mathrm{keV}$ energy range (left, 300 s bins), and from the two \emph{XMM-Newton} analyzed observations in the $0.3-10~\mathrm{keV}$ energy range (right, 600 s bins, observation 0694640501 and 0694641201 in black and grey, respectively).}
\end{figure*}

We extracted the spectra of the EPIC-pn camera in the $0.3-10~\mathrm{keV}$ energy range following the standard procedure described in the \emph{XMM-Newton} software analysis guide\footnote{\footnotesize{\href{ftp://legacy.gsfc.nasa.gov/xmm/doc/xmm_abc_guide.pdf}{The XMM-Newton ABC Guide: An Introduction to XMM-Newton Data Analysis}}} for the observations 0694640401, 0694640501, and 0694641201, using the SAS software release xmmsas\_20131209\_1901-13.5.0. The lightcurves and spectra of \object{1E1743.1-2843} were extracted from annular extraction regions excluding zones where the number of counts exceeded 800 to avoid pileup if present (see~\figurename~\ref{xmmregion}). The lightcurves are shown in \figurename~\ref{lc}. 

EPIC-pn data below $0.3~\mathrm{keV}$ are mostly related to artifacts and noise and were excluded \citep{xmmguide}, and all the bins were grouped to reach at least 30 counts. The mean count rate for the two observations is $2.62\pm0.01~\mathrm{cts/s}$. 

\subsection{\emph{Timing analysis} }

\begin{figure}[h]
 \centering
 \includegraphics[width=\linewidth]{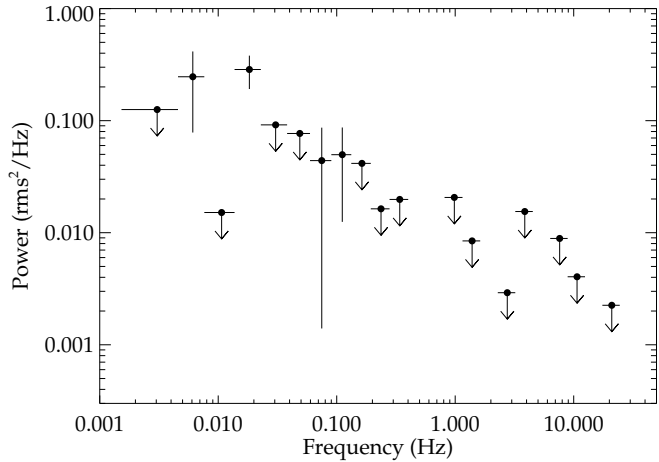} 
 \vspace{4ex}
 \caption{\label{fig_lombxmm}Power spectrum of the \protect\emph{NuSTAR} observation of \protect\object{1E1743.1-2843}, with indication of a low frequency break at frequencies $>0.02$~~Hz. Upper limits are 1$\sigma$.}
\end{figure}

A lightcurve of the \emph{NuSTAR} observation with 10~ms time bin resolution was extracted for FPMA in the 
energy range 3--60~keV\footnote{\footnotesize{{ The low energy range is different from the one used in 
the spectral analysis section, in the attempt to maximize the SNR}}} using \emph{nuproducts}. Timing data 
were corrected for deadtime and for arrival times at the Solar System barycenter, using the JPL 2000 
ephemeris \citep{ephemeris} (for this purpose, we used the barycorr tool in the HEASOFT 6.16 distribution). We then calculated the power spectra on different contiguous time intervals and averaged them into a 
total spectrum. Each single spectrum was built using intervals of 32768 bins, and the total spectrum was built 
averaging 87 intervals in a single frame. This was finally rebinned in frequency channels for more statistics.
An offset constant term was subtracted from the total spectrum to remove the Poisson noise level and 
compensate for residual effects of the deadtime correction (this term was evaluated as the average power in the frequency interval 10--49~Hz). 
The resulting power spectrum is shown in Figure \ref{fig_lombxmm}. The statistics is too poor 
to model the power spectrum with a multi-parameter function (such as a Lorentzian). Signal is detected only up
to frequencies of $\sim0.1$~Hz, with some hint of complex structure, like the presence of a possible break at frequencies $>0.02$~Hz. 

We also performed a Lomb-Scargle periodogram of the same frequency interval as the power spectrum, and found no significant
signal. The power spectrum and Lomb period analysis were also performed for the two \emph{XMM-Newton} observations, also yielding no 
positive detection of periodicities.

\subsection{\emph{Spectral analysis}} 

We analyzed the \emph{NuSTAR} and the \emph{XMM-Newton} datasets using XSPEC \citep{Arnaud} version 12.8.0, using the {\tt TBabs} with abundances set as in \citet{Wilms} and cross sections set as in \citet{Verner} to model the effect of X-ray absorption. We performed the fit allowing the normalizations among the three different observations to vary. These normalizations relative to the different \emph{XMM-Newton} datasets remained constant for every model, and were found to be $C_{1}=0.99\pm0.01,~C_{2}=1.00\pm0.01$ for observations 0694641201 and 0694640501, respectively. 

We used the following four models to figure out the nature of the source, the first three of which were used to test the origin of the low-energy emission:

\begin{itemize} 
\item Model 1: a black body ({\tt bbodyrad}) and a power-law with a high-energy cutoff (\hbox{\tt power-law $\times$ highEcut}). This is a typical single-component spectral model, and assumes the source is a neutron star binary.
\item Model 2: a disk black body ({\tt diskbb}) plus a \hbox{\tt power-law $\times$ highEcut}. This is also a typical single-component spectral model, but assumes the compact source is a black hole binary.
\item Model 3: a disk black body ({\tt diskbb}), a black body ({\tt bbodyrad}), and the \hbox{\tt power-law $\times$ highEcut}. This model probes deeper into the neutron star hypothesis, adding an accretion disk component to model 1.
\item Model 4: a disk black body ({\tt diskbb}), a black body ({\tt bbodyrad}), and the {\tt compTT} Comptonization model. In this model the powerlaw in model 3 is replaced by a more physical model.
\end{itemize} 

To model the emission from the accretion disk we used the multi-color disk black body {\tt diskbb} model mentioned before \citep{diskBB1, diskBB2}.

The fit results shown in \tablename~\ref{allfit} indicate that the source emission is
mainly contributed by a prominent blackbody component at $\sim2$~keV plus a
high energy continuum which can be described equally well by an
empirical law (\hbox{\tt power-law $\times$ highEcut}) or by a thermal Comptonization
({\tt compTT}) component. As can be seen from \tablename~\ref{allfit} all the models provide a good fit of the data. In the following we will assume as baseline model 3, since it provides the lowest $\chi^{2}$ value, and a reasonable physical interpretation of the data, as discussed in section~\ref{sec:results}.

Adding another absorption component with partial covering ({\tt pcfabs}) to model 1 did not improve the fit results, therefore we conclude that the source is not partially obscured.
Also, the spectroscopic data do not need the addition of further high energy components such as, for instance, reflection, as we verified by adding the {\tt coplrefl} \citep{Ballantine2012} reflection component to model 3.

We then addressed the high-energy component taking as a baseline model 3 and replacing the power-law plus cutoff component with a more physical model, specifically {\tt compTT} \citep{Titarchuk}, which describes the Comptonization of soft photons in a thermal plasma of high-energy electrons above the accreting source.
In the resulting model 4 the {\tt compTT} component gives a worse fit than the power-law with an exponential cutoff (model 3). However, model 4 still yields a better fit than models 1 and 2, and has the advantage of allowing a more physical interpretation of the source spectrum than the empirical model 3. 
Furthermore, we tested if a single thermal component and a Comptonized one could fit the data, and replaced the power-law in model 1 with the {\tt compTT} model. However this resulted in a worse fit $(\chi^{2}_{red}\sim1.07)$.
Finally, if we test model 4 for the presence of the fluorescence line of neutral iron, $\chi^{2}_{red}$ rises to 1.2. Therefore the presence of the iron $K_{\alpha}$ line is also not required. From model 4 we derive an upper limit on the iron $K_{\alpha}$ equivalent width (EW) of 4.9 eV.

\begin{figure*}
 \label{allspec}
 \begin{minipage}[b]{0.5\linewidth}
 \centering
 \includegraphics[width=1\linewidth]{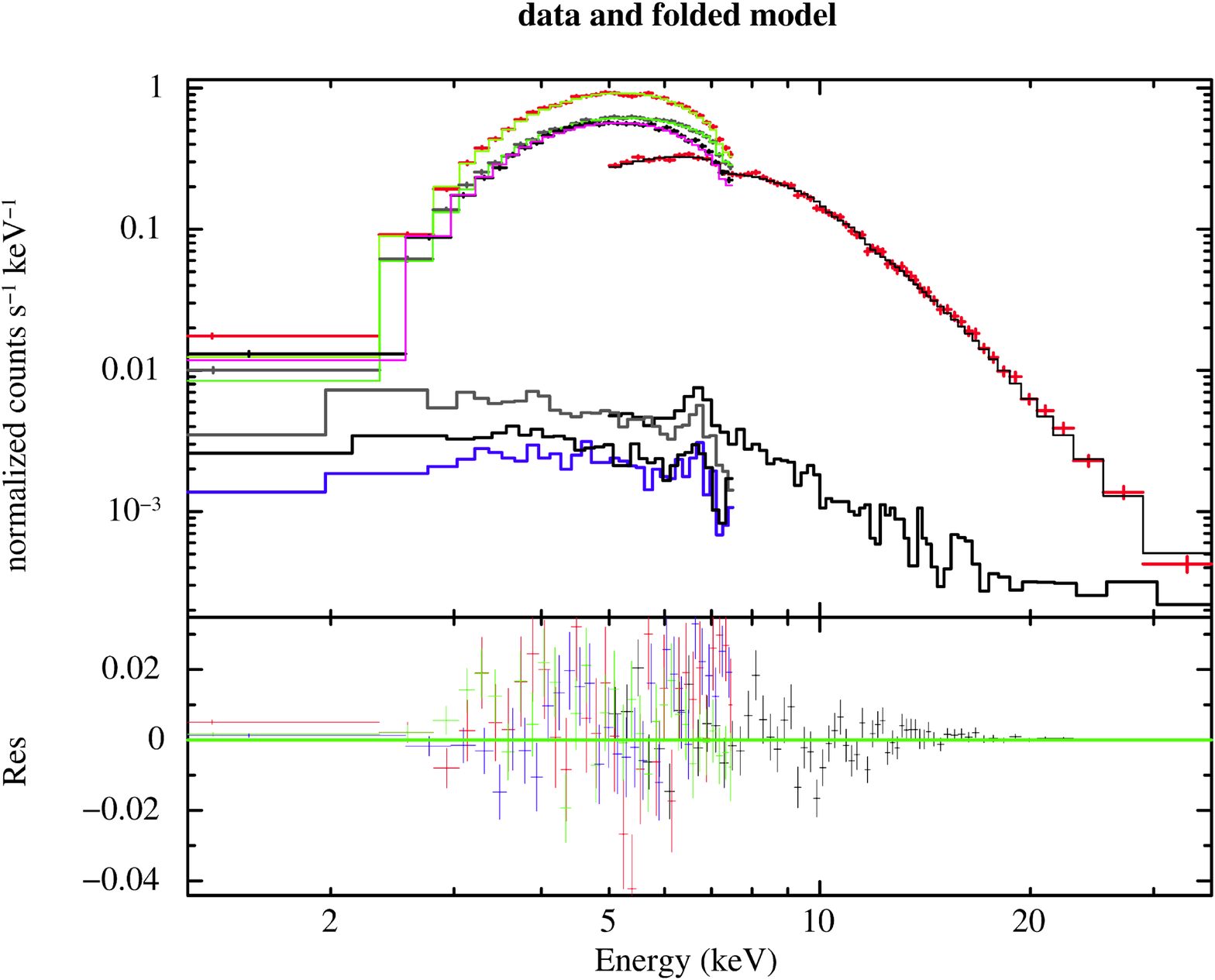} 
 \vspace{4ex}
 \end{minipage}
 \begin{minipage}[b]{0.5\linewidth}
 \centering
 \includegraphics[width=1\linewidth]{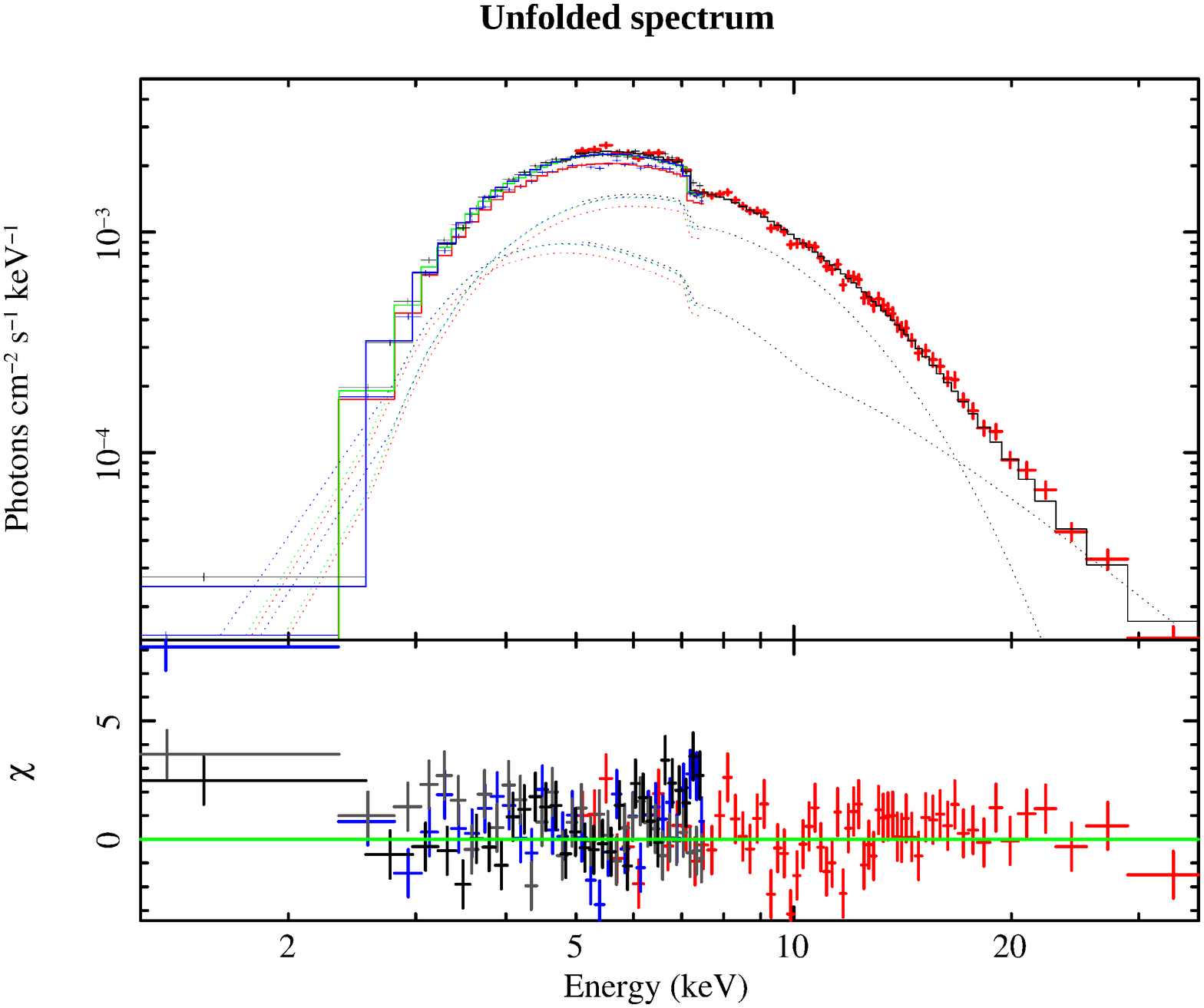} 
 \vspace{4ex}
 \end{minipage}
 \caption{Left - folded spectra, backgrounds and residuals with respect to model 4 for all the spectra analyzed. Right - the analyzed spectra, unfolded through model 4. The plots are rebinned in groups of 30 channels for display purposes.}
\end{figure*}

Summarizing, model 3 and model 4 give the best results. Model 3 has the lower reduced $\chi^{2}$ value among the two, though the {\tt compTT} model allows the determination of the physical parameters of the source.
The resulting spectra for model 4 are shown in \figurename~\ref{allspec}, and the results of the overall fits in are reported in \tablename~\ref{allfit}. 

 \begin{table*}
\begin{center}
\small
\caption{Fit results of the \protect\emph{NuSTAR} and \protect\emph{XMM-Newton} observations.}
 \label{allfit}
\begin{tabular}{ccccccc}
\tableline\\ [-2.0ex]
XSPEC Model & Parameter & Units & Model 1 & Model 2 &Model 3 & Model 4 \\
\tableline\\ [-2.0ex]
TBabs & $N_{H}$&$10^{22}~\mathrm{cm^{-2}}$	& $14.8_{-0.8}^{+0.9}$ 		& $16.6\pm0.5$ 		& $16.5_{-1.3}^{+1.0}$ 		& $15.3_{-1.1}^{+0.8}$ \\
bbodyrad & $kT$&$\mathrm{keV}$ 			& $1.8\pm0.1$		 	& 			& $1.8_{-0.2}^{+0.1}$		& $1.83\pm0.05$\\
bbodyrad & norm\tablenotemark{a}&		& $1.3_{-0.3}^{+0.5}$ 		& 			& $1.0\pm0.3$ 			& $1.7_{-0.2}^{+0.4}$\\
diskbb & $kT_{in}$&$\mathrm{keV} $		& 				& $2.4_{-0.3}^{+0.2}$	& $0.12_{-0.02}^{+0.04}$ 	& $0.12_{-0.03}^{+0.04}$\\
diskbb & norm\tablenotemark{b}&			& 				& $0.19_{-0.08}^{+0.1}$	&$-$				& $-$\\
power-law & $\Gamma$ &				& $1.0_{-0.5}^{+0.3}$ 		& $0.9_{-0.4}^{+0.3}$	& $1.3_{-0.3}^{+0.2}$ 		& \\
power-law & norm\tablenotemark{c}&		&$0.009_{-0.006}^{+0.007}$	&$0.01_{-0.005}^{+0.01}$&$0.02\pm0.01$			& \\
highecut & cutoffE&$\mathrm{keV}$ 		& $6.6\pm0.4$ 			& $6.9\pm0.30$ 		& $7.2_{-0.7}^{+0.4}$ 		& \\
highecut & foldE &$\mathrm{keV}$ 		& $8.1_{-1.5}^{+1.7}$ 		& $6.3_{-1.3}^{+0.5}$	& $9.0_{-1.3}^{+1.8}$ 		& \\
compTT & $T_{0}$&$\mathrm{keV}$			&				& 			& 				& $0.014_{-2.9}^{+8.4}$\\ 
compTT & $kT$&$\mathrm{keV}$ 			&				& 			& 				& $4.6_{-0.4}^{+0.6}$\\ 
compTT & $\tau_{p}$ &				&				& 			& 				& $6.2_{-1.4}^{+6.5}$\\
compTT & norm\tablenotemark{d}&			& 				& 			&	 			& $0.031_{-0.013}^{+93}$\\
$\frac{\chi^{2}}{d.o.f.}$ &&			&$\frac{2798.3}{2637}$ 		& $\frac{2835.4}{2637}$	& $\frac{2766.4}{2635}$ 	& $\frac{2785.0}{2635}$\\
reduced $\chi^{2}$ 	&&			&1.061				&1.075			&1.050				&1.057 \\
\tableline\\ [-2.0ex]
\end{tabular}
\tablecomments{The errors are expressed with 90\% confidence. Parameters not reported were not constrained by the fit.\\
$^a$ $\sfrac{R^{2}_{km}}{D^{2}_{10}}$, where $R_{km}$ is the source radius in km, and $D_{10}$ is the distance to the source in units of 10 kpc.\\
$^b$ $(\sfrac{R_{in [km]}}{D_{10}})^{2}cos\theta$, where $R_{in [km]}$ is the apparent disk radius, and $\theta$ is the angle of the disk ($\theta=0$ is face-on)\\
$^c$ The powerlaw component normalization in units of $\mathrm{photons~keV^{-1}~cm^{-2}~s^{-1}}$\\
$^d$ The thermal Comptonization component normalization in units of $\mathrm{photons~keV^{-1}~cm^{-2}~s^{-1}}$\\
}
 \end{center}
 \end{table*}

\section{Discussion} \label{sec:results}

In this work we have reported on a broad band ($0.3-40~\mathrm{keV}$) spectral analysis of \object{1E1743.1-2843}, observed with \emph{NuSTAR} and \emph{XMM-Newton}. A similar analysis had previously been performed using \emph{INTEGRAL} data \citep{delsanto2006}, but the improved energy resolution and sensitivity of \emph{NuSTAR} allows us to better constrain the hard X-ray continuum, identifying the presence of Comptonization and of a cutoff in the high-energy emission for the first time. This allowed to use more sophisticated models compared to previous works \citep{delsanto2006,porquet2003,cremonesi1999}. Below we summarize the results obtained and discuss the ones that support either the LMXB or HMXB nature of the system.

\begin{figure}[h]
 \centering
 \includegraphics[width=\linewidth]{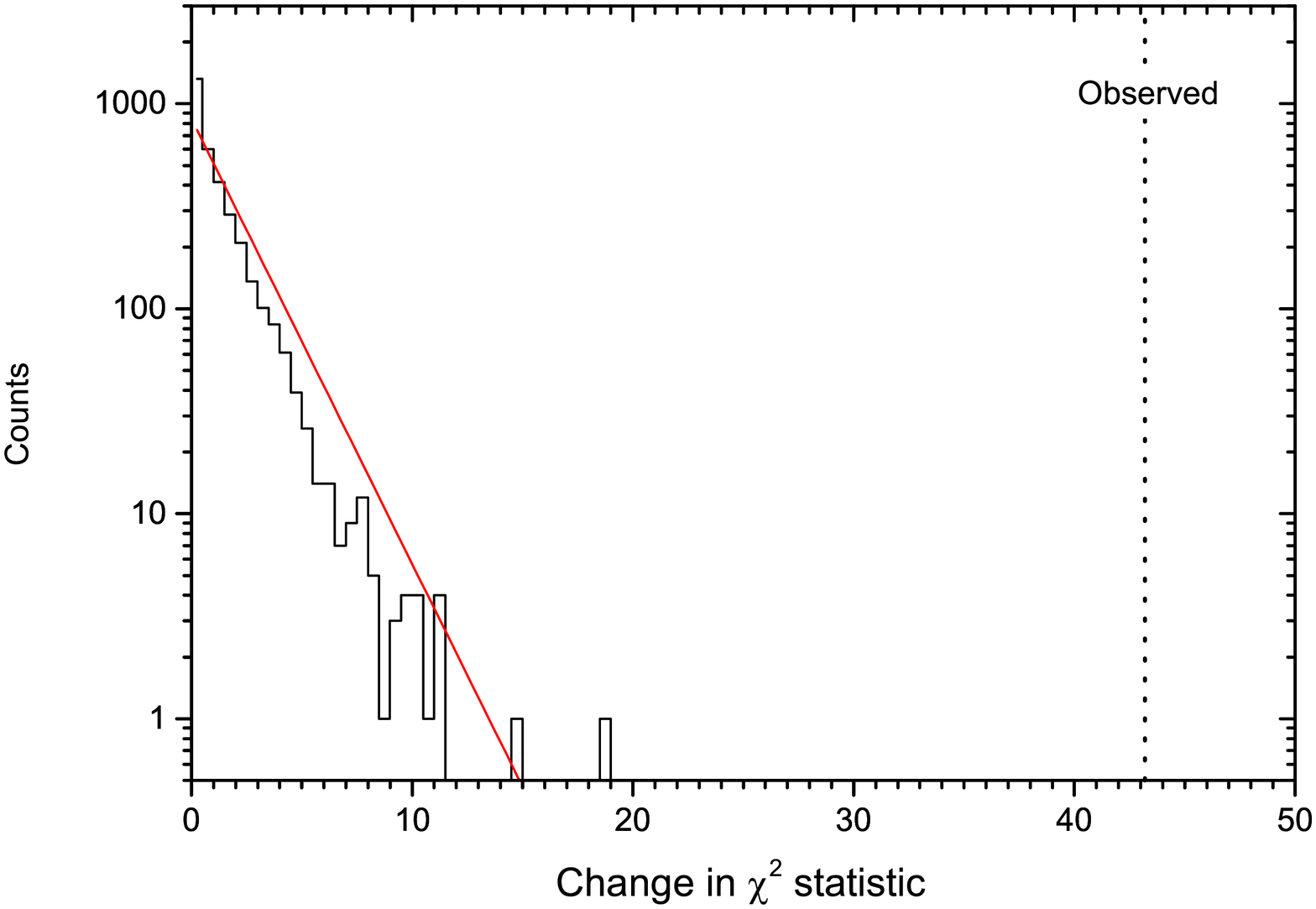} 
 \vspace{4ex}
 \caption{\label{deltachi} Results of Monte Carlo simulations for testing the presence of the disk component. We simulated the spectra from model 3 without (our null-hypothesis) and with the disk component. The histogram shows the $\Delta\chi^2$ obtained in 10000 simulations (in black), together with the $\Delta\chi^2$ distribution expected from the addition of 2 free parameters (red line). The vertical dashed line is the $\Delta\chi^2$ value obtained from the actual data (43.19).}
\end{figure}

Regarding the low-energy emission of the source, the presence of the kT=1.8~keV black body is a strong indication of a neutron star nature of the compact object. The black body radius is $\sim1$~km if we assume the source to be located in the Galactic Center, at $d\sim8.8~\mathrm{kpc}$. This is not compatible with emission from a boundary layer near the neutron star surface ($R_{BL}-R_{NS}\sim2$~km for low \.{M}, \citealp{Popham2001}), but is consistent with emission from a restricted region of the neutron star surface (i.e., from an equatorial belt in the orbital plane, or with magnetically driven accretion onto polar caps). In the latter case the magnetic field strength of the compact source should be $>10^{9}$~G.
There is possibly a soft excess which could be interpreted as weak emission from an accretion disk ($kT\sim0.1~\mathrm{keV}$). However, the \emph{XMM-Newton} data were unable to constrain the black body normalization due to the high hydrogen column absorption ($N_{H}\sim1.6\times10^{23}~\mathrm{cm^{-2}}$). In the HMXB scenario this soft excess could also be interpreted as a blend of emission lines, as thermal emission from the neutron star surface, or Thomson scattering of the hot spot emission by the accreting material \citep{vandermer}. The reliability of the soft excess detection is worth discussing, because the $\chi^{2}$ improvement it provides is moderate. 
The use of the F-test could result in unreliable results \citep{protassov}, so we performed a series of Monte Carlo simulations with the \emph{simftest} routine to confirm the presence of such component, following the approach described by \citet{bhalerao}. The highest $\Delta\chi^2$ obtained in 10000 simulations is 18.7, significantly lower than $\Delta\chi^2=43.2$ obtained by the real data (see \figurename~\ref{deltachi}). We estimate that $>10^9$ simulations would be required to get $\Delta\chi^2>40$, corresponding to a significance higher than $6\sigma$.
Nevertheless, even if the presence of the soft excess is statistically preferred, we cannot exclude that the low energy spectrum is affected by systematics, which could arise, e.g., in adding two exposures taken at different epochs. We regard this as a result that needs to be confirmed by better quality low energy data from future observations of this source.

The presence of a neutral iron $K_{\alpha}$ line at 6.4 keV is not statistically required, with an upper limit on the EW of 4.9 eV from model 4 (6.7 eV from model 3). This value is compatible with the expected properties of LMXB \citep{asai2000} and radio-quiet quasars \citep{george2000}, as already pointed out by Porquet et al. (2003). This weakens the HMXB hypothesis for this source, since the iron line is usually detected in such systems. On the other hand, lack of the Fe $K_{\alpha}$ line in LMXBs is very common, and is usually associated with the Baldwin effect: the high luminosity of the X-ray source increases the degree of iron ionization \citep{felack}. 

The high-energy emission is characterized by a power-law ($\Gamma \sim 1.3$) with a high-energy cutoff ($E_{cut}\sim7.2~\mathrm{keV}$), identified for the first time thanks to the \emph{NuSTAR} hard X-Ray sensitivity. The \emph{NuSTAR} data also unambiguously identified the presence of a Comptonization component, induced by an electron population with temperature of a few keV ($kT_{0}\sim4.6~\mathrm{keV}$, in accordance with the exponential cutoff value found), indicating a large viewing angle (see \citealp{angle}). 
Even if X-rays are produced near the neutron star at energies of several tens of keV, the observed X-rays will be shifted to lower energy due to Comptonization by the more distant low-energy plasma with a temperature several keV.
This Comptonization component has a strong interplay with the 1.8 keV black body up to $\sim20~\mathrm{keV}$, since the total hard X-ray spectrum cannot be described as a single power-law. We tested for the presence of a reflection component, and found that it is not required to explain the data.
A fit with a power-law model without breaks brought higher values of the photon index ($\Gamma=2.3^{+0.08}_{-0.09}$), compatible with previous results \citep{delsanto2006,porquet2003,cremonesi1999}, but also produced a higher value of the $\chi^{2}/d.o.f.=2880.79/2639=1.09$ compared to the one obtained with the high-energy cutoff ($\chi^{2}_{red}=1.05$; see \tablename~\ref{allfit}), indicating that using the former description for the higher-energy data is marginally justified at best.

\begin{figure}[h] 
 \label{ukidss}
 \centering
 \includegraphics[width=\linewidth]{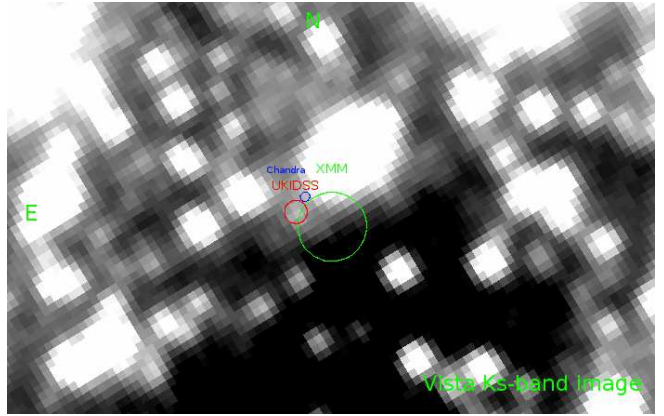} 
 \vspace{4ex}
\caption{The $Ks$-band image of \protect\object{1E1743.1-2843} obtained with Vista. 
The \protect\emph{Chandra} and \protect\emph{XMM-Newton} positions of \protect\object{1E1743.1-2843} are shown, together with the position of \protect\object{UGPS~J174621.12284343.3} as reported in the UKIDSS catalogue (coordinates: R.A.~=~17:46:21.13, $\delta$~=~-28:43:43.17~J2000). The \emph{Chandra} error circle excludes the possibility of this source being the counterpart.
 }
\end{figure}

The source luminosity is $L_{2-10~\mathrm{keV}}\sim1.5\times10^{36}~d^{2}_{10~\mathrm{kpc}}~\mathrm{erg~s^{-1}}$, so if we assume that the source is located in the Galactic Center we obtain $L_{2-10~\mathrm{keV}}\sim10^{36}~\mathrm{erg~s^{-1}}$, within the typical range of luminosity for X-ray bursters \citep{cremonesi1999,bursters}. To investigate further, we have extracted the $Ks$-band image of the Vista survey ``VISTA Variables in the Via Lactea (VVV)'' \citep{vistasurvey} (see \figurename~\ref{ukidss}). Both the \emph{XMM-Newton} and \emph{Chandra} positions of \object{1E1743.1-2843} are visible in the image, as well as the position of the source UGPS~J174621.12284343.3 as reported in the UKIDSS catalogue \citep{ukidsscatalog}. 
The nominal accuracy of UKIDSS is $\sim0.1~\mathrm{arcsec}$, but to account for source confusion in this crowded region we used a larger $0.3~\mathrm{arcsec}$ positional error, as suggested by \citet{ukidsscatalogunc}. We should consider whether this source might be the counterpart. In the UKIDSS catalog, this source does not have a $K$-band magnitude listed, but it has $H=16.82\pm0.09$ and $J=18.39\pm0.08$. Assuming the $0.21''$ uncertainty on the \emph{Chandra} position \citep{chandrapos}, then this source is excluded, as can be seen in the figure. Also, given the extreme IR source crowding in the GC, the fraction of true-to-candidate counterparts for hard \emph{Chandra} sources is very low (see, for instance \citealp{DeWitt2010}). Furthermore, since the \emph{NuSTAR/XMM-Newton} column densities are consistent with the overall Galactic Center value, it not likely that the companion star is hidden by a cloud either. 
 However, we cannot exclude a high mass companion solely on the basis of the absence of a bright IR source. In fact, a B-type star placed at the distance of 8.5 kpc could have IR magnitudes fainter than the limiting H-band magnitude of the vista surveys of the Galactic Center \citep{vhs}, assuming an IR absorption corresponding to the hydrogen column density we measured for \object{1E1743.1-2843} (see \tablename~\ref{allfit}).

In the LMXB hypothesis the fact that not a single burst has been observed in over 20 years suggests that the source is a rare burster \citep{lilzand2004}. In principle, the lack of bursts could be explained if the source is located outside the Galaxy. Specifically, to reach $L=2.9\times10^{37}~\mathrm{erg~s^{-1}}$, corresponding to the stable burning of accreted material, \object{1E1743.1-2843} would need to be placed at $d\sim40~\mathrm{kpc}$. However, a study of the stability conditions for accreting objects \citep{narayan2002} indicates that, for a neutron star with a surface temperature of$\sim2$~keV, there is a luminosity range where stable accretion is possible, between the instability regions where He and H bursts are triggered. This ``stability strip'' corresponds roughly\footnote{\footnotesize{{ Exact boundaries depend weakly on the neutron star radius.}}} to luminosities $1.3\times10^{36}<L<6.5\times10^{36}~\mathrm{erg~s^{-1}}$. This implies that if the
 source is located between the Galactic Center and $\sim16~\mathrm{kpc}$ the lack of bursts is to be expected, as well as the lack of the detection of an IR companion.
	
\citet{church2014} recently proposed a unified model for the LMXB sources. The model assumes the presence of an extended accretion disk corona (ADC). For $L>1-2\times10^{37}~erg~s^{-1}$ the ADC is in thermal equilibrium with the neutron star surface, giving rise to a Comptonized spectrum with $E_{cut}\sim~6~\mathrm{keV}$, which corresponds roughly to three times the actual temperature of the electrons (in the assumption of high optical depth). For lower luminosities, the Comptonization becomes inefficient in cooling the corona, the thermal equilibrium assumption breaks down and, as a consequence, the extended ADC heats up to several tens of keV. In this scenario the $\sim1$~km black body radii in LMXBs are explained by an emitting region in the shape of an equatorial strip in the orbital plane with a half height $h\sim100$~m.
This scenario is described by a black body plus a cutoff power-law (our model~1). The authors also note that compTT is not consistent with the evidence for an extended corona. The outer regions of the accretion disk ($kT\sim0.1$~keV), which provide the seed photons for the Compton scattering, are not expected to be detected.
Our results for the black body temperature ($kT\sim1.8$ keV) and radius ($r\sim1$~km), for the $E_{cut}~\sim~3kT_{e^{-}}\sim6.6~\mathrm{keV}$ of the Comptonized component, and for the ratio of total to black body luminosity for \object{1E1743.1-2843} fit within the expected ranges for the scenario they depict as the Banana state of the Atoll sources. This interpretation, however, implies $L>2\times10^{37}~\mathrm{erg~s^{-1}}$ to explain the spectral cutoff value and the lack of bursting activity. This would require a distance of a few tens of kpc to account for the observed flux, which could be explained if \object{1E1743.1-2843} were in the Sagittarius dwarf elliptical galaxy (SAGDEG), one of the small dwarf spheroidal galaxies that orbits the Milky Way. SAGDEG is currently behind the GC at distance $d\sim26~$kpc \citep{sagdeg}, and is in an advanced state of destruction due to tidal interactions with the Galaxy. Therefore a fraction of the stars that composed this dwarf galaxy have likely scattered to even greater distances.

In the HMXB scenario, the presence of a strong magnetic field ($B\sim10^{12-13}~\mathrm{G}$) suppresses the propagation of the bursts across the neutron star surface \citep{gilfanov&sunyaev}. However the lack of eclipses, Fe $K_{\alpha}$ line, and pulsations in the light curve, as well as the missing detection of a companion star, makes the HMXB hypothesis less favored, even though the value of the spectral cutoff ($9\pm1.8~\mathrm{keV}$) is compatible with the one expected from HMXBs ($10-20~\mathrm{keV}$).

We conclude that while an HMXB framework leaves several unexplained features, interpreting \object{1E1743.1-2843} as a NS-LMXB scenario is more consistent, implying a peculiar but not a unique object. In this case the source could be located at a distance $9<d<16$ kpc, between the two instability luminosity intervals where He and H bursts are triggered or, if we rely on the unified model proposed by \citet{church2014}, at a distance $d>36~$kpc, corresponding to the stable burning of accreted material. 

\acknowledgments

This work was supported under NASA Contract No. NNG08FD60C, and made use of data from the \emph{NuSTAR} mission, a project led by the California Institute of Technology, managed by the Jet Propulsion Laboratory, and funded by the National Aeronautics and Space Administration. We thank the \emph{NuSTAR} Operations, Software and Calibration teams for support with the execution and analysis of these observations. This research has made use of the \emph{NuSTAR} Data Analysis Software (NuSTARDAS) jointly developed by the ASI Science Data Center (ASDC, Italy) and the California Institute of Technology (USA). The Italian authors acknowledge
the Italian Space Agency (ASI) for financial support by ASI/INAF grant I/037/12/0. RK acknowledges support from Russian Science Foundation (grant 14-22-00271).

\end{document}